\documentclass[aps,pre,twocolumn,amsmath,amssymb,superscriptaddress,floatfix,showpacs]{revtex4}
\usepackage{graphicx}
\usepackage{bm}
\input epsf.sty

\begin{document}
\title{Changing growth conditions during surface growth}
\author{Yen-Liang Chou}
\author{Michel Pleimling}
\author{R. K. P. Zia}
\affiliation{Department of Physics, Virginia Polytechnic Institute and State University, Blacksburg, Virginia 24061-0435, USA}

\begin{abstract}
Motivated by a series of experiments that revealed a temperature dependence
of the dynamic scaling regime of growing surfaces, we investigate
theoretically how a nonequilibrium growth process reacts to a sudden change
of system parameters. We discuss quenches between correlated regimes through
exact expressions derived from the stochastic Edwards-Wilkinson equation with a variable
diffusion constant.
Our study reveals that a sudden change of the diffusion constant
leads to remarkable changes in the surface roughness. Different dynamic
regimes, characterized by a power-law or by an exponential relaxation, are
identified, and a dynamic phase diagram is constructed. We conclude that growth
processes provide one of the rare instances where quenches between correlated regimes
yield a power-law relaxation.
\end{abstract}


\date{\today }
\maketitle

\affiliation{Department of Physics, Virginia Polytechnic Institute and State
University, Blacksburg, Virginia 24061-0435, USA}

\section{Introduction}

Because of their ubiquity in nature, nonequilibrium growth processes have
been the subject of numerous studies during the last two decades \cite
{Mea93,Hal95,Bar95}. Remarkably, these studies revealed very general
properties of growing interfaces that are encountered in a large variety of
growth processes, ranging from crystal growth to tumor growth. In this
context, due to its obvious technological relevance, thin film growth has
been one of the major research thrusts.

{}From the theoretical point of view, many important insights into the
universal aspects of nonequilibrium growth processes have been obtained
through the study of stochastic differential equations and of simple model
systems \cite{Kru97,Kru95}. One of the simplest approach is due to Edwards
and Wilkinson \cite{1982SE} who described the surface growth due to particle
sedimentation by $\partial h({\bf x},t)/\partial t=\nu \nabla ^2h(%
{\bf x},t)+u$, where $h({\bf x},t)$ is the surface height at
time $t$ at a site ${\bf x}$ of a $d$-dimensional substrate (of area 
${\cal A}$) and $u$ represents a constant flux of deposition. The physical
origin of the ``diffusion constant'' $\nu $ can be traced to the surface
tension as well as $T$, the temperature of the substrate. When noise is
added to this and the process is viewed from a co-moving frame of the steady
state (i.e., $h({\bf x},t)-ut\rightarrow h({\bf x},t)$), we arrive at the
stochastic Edwards-Wilkinson (EW) equation 
\begin{equation}
\frac{\partial h({\bf x},t)}{\partial t}=\nu \nabla ^2h({\bf x},t)+\eta (%
{\bf x},t)  \label{eqEW}
\end{equation}
where $\eta ({\bf x},t)$ is a Gaussian white noise with zero mean and
covariance $\left\langle \eta ({\bf x},t)\eta ({\bf y},s)\right\rangle
=D\delta ^d({\bf x}-{\bf y})\delta (t-s)$. 
A microscopic realization of the Edwards-Wilkinson equation was soon
proposed by Family \cite{1986FF,Bar95} (see also Ref. [\onlinecite{Rot06}]
for a recent more in-depth comparison of that model with the EW equation).
In the random deposition with surface relaxation (RDSR) process particles
drop from randomly chosen sites over the surface. Instead of sticking to the
surface at the point of impact, the particles are allowed to explore the
nearest vicinity of that point and are finally incorporated into the surface
at the neighboring site with lowest height. This diffusion step smoothes the
surface and yields correlated growth.

The roughness of a growing surface is characterized by the time dependent
mean interface width 
\begin{equation}
W(t)=\sqrt{\langle \left( h-\overline{h}\right) ^2\rangle }
\end{equation}
where $\overline{h}(t)\equiv \int h({\bf x},t)d^d{\bf x}/{\cal A}$ is the
mean surface height at time $t$. In many instances the mean interface width
(after an initial regime of random surface growth when starting from a flat
initial condition) displays a power-law dependence on time, $W(t)\sim
t^\beta $, before saturating at a value $W\sim L^\alpha $ where $L$ is the
size of the system. The growth exponent $\beta $ and the roughness exponent $%
\alpha $ are universal quantities that characterize large classes of growth
systems belonging to the same growth universality class. Thus for the
Edwards-Wilkinson universality class one finds for a one-dimensional
substrate $\beta =1/4$ and $\alpha =1/2$. Other universality classes have
also been found \cite{Kru97,Kru95}, some of which, as for example the
Kardar-Parisi-Zhang (KPZ) \cite{Kar86} and the conserved KPZ universality
classes \cite{Wol90,Sar91}, are of direct relevance for thin film growth.

The morphology of growing structures can depend crucially on experimental
parameters as for example the flux of deposited particles or the substrate
temperature \cite{Jen99}. Different experimental groups have reported a
temperature dependence of the roughness of growing or sputtered surfaces,
and temperature dependent values of the growth exponents have been found in
some systems \cite
{Ern94,Yan94a,Yan94b,Ell96,Smi97a,Smi97b,Sto00,Kal01,Bot01,Els04,Fer06}.
Systems for which this has been observed include homoepitaxial growth on
Cu(001) \cite{Ern94,Bot01}, Ag(100) \cite{Ell96,Sto00}, and Ag(111) \cite
{Ell96}, amorphous thin-film growth \cite{Els04}, growth of CdTe
polycrystalline films \cite{Fer06}, molecular-beam epitaxy growth of
Si/Si(111) \cite{Yan94b}, as well as ion-sputtered Si(111) \cite{Yan94a},
Ge(001) \cite{Smi97a,Smi97b} or Pt(111) \cite{Kal01} surfaces. The
dependence of the roughness on temperature can be rather involved, yielding
different types of behavior for different systems, depending on how
diffusion takes place and whether additional smoothening mechanisms are
present. In many instances one observes that growth becomes increasingly
rougher with decreasing temperature, yielding an increase of the value of $%
\beta$. On the other hand, some experiments \cite{Els04,Fer06} revealed an
increase of the global roughness with temperature.

The observation of a transition between different dynamic scaling regimes
due to a change of experimental conditions is very intriguing and raises the
question how a growing surface evolves from one regime to another after a
sudden change of, for example, the substrate temperature. 
We are not aware of any experimental study of growth processes where this
kind of protocol has been implemented. However, we expect that the intriguing
results reported here will motivate future experimental studies along similar
lines.
In this paper we discuss exact results derived from the EW
equation with a variable diffusion constant, 
which allows us to investigate systematically the changes in the
surface roughness in case growth conditions are changed {\it during} the growth
process. Interestingly, different dynamic regimes are encountered, some of
which are characterized by a power-law relaxation. Here, we do not have a
specific experimental system in mind. Instead, our interest is broader,
namely, to understand universal responses of a growing surface to sudden
changes of experimental parameters through the study of simple models.

It is to be noted that our study is complementary to an earlier investigation
due to Majaniemi, Ala-Nissila, and Krug \cite{Maj96}. Similarly to our work, these
authors studied the impact of a change of growth conditions on processes described
by linear growth equations. Whereas we discuss in the following the effects of a variable diffusion 
constant, Majaniemi et al. analyzed how the growing surface reacts to a change
of the noise in the system.

There is also a second, theoretical motivation for our study.
Sudden changes of external conditions, as for example due to a temperature quench,
have been investigated in recent years in a large variety of
systems, ranging from magnetic systems to glasses \cite{Cug,Ple08,Hen09}. In the
most common setting, a system initially prepared in some equilibrium state is suddenly brought 
out of equilibrium through a temperature quench. If the system is characterized
by slow, non-exponential relaxation (as it is the case for a ferromagnet quenched to or below the
critical point), interesting nonequilibrium processes and aging phenomena are observed \cite{Hen09}.
Some studies also focused on slow relaxation encountered in up-quenches in which a magnetic system 
initially in the ground state is
quenched to the critical point \cite{Gar05,Pau07}. Interestingly, however, slow relaxation is usually not
observed when both the initial and final temperatures are below the critical point. Indeed, 
for models with a discrete global symmetry such as Ising or Potts models a competing ordered state
can not be reached if the starting point is close to one of the minima of the equilibrium
free energy. Consequently, non-exponential relaxation is only encountered in this type of quench
if the system has a continuous symmetry, as it is for example the case of the $XY$ model \cite{Ber01,Pic04}.
As we will show in this paper, surface growth processes constitute an interesting class of systems where
a change of parameters yields a transition from one correlated state to another characterized by a power-law relaxation.


The paper is organized in the following way. In Section II we discuss the dependence of 
the surface width, derived from the EW equation, on the value of
the diffusion constant.
Section III is devoted to the study of the time evolution of the surface
roughness following a sudden change of the diffusion constant.
We thereby identify different dynamic regimes and present a
dynamic phase diagram that summarizes the possible responses of the growing
interface. Finally, we end with a summary
and outlook in Section IV.

\section{Interface width}

In order to study the effect of a change of external conditions on simple growth processes
we consider the Edwards-Wilkinson equation (\ref{eqEW}) with a variable diffusion
constant $\nu$ (a microscopic realization of this process has recently
been discussed in \cite{Cho09}). When describing a deposition or growth process by this equation,
one implicitly assumes that $\nu$ and the noise amplitude $D$ depend on the experimental parameters, 
as for example the temperature $T$. In the following we will not need to know the
explicit dependence of $\nu$ and $D$ on these parameters (which would be system dependent), and study
how the interface width changes when changing the value of $\nu$. The reaction of the growing surface
to a change of the noise has been studied in \cite{Maj96}.

\begin{figure}[h]
\centerline{\epsfxsize=3.40in\ \epsfbox{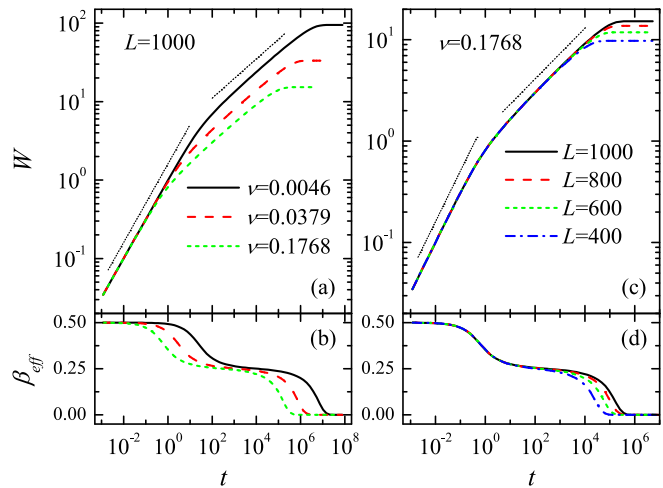}}
\caption{(a) Log-log plot of the surface width vs time for a system of size 
$L=1000$ and different diffusion constants. The dotted lines have the slopes $1/2$
and $1/4$ expected in the random deposition and EW regimes, respectively.
The locations of both crossover points depend on the diffusion constant. The data
are obtained from the exact solution of the EW stochastic equation.
(b) Time evolution of the effective exponent (\ref{eqBeta}) for the data shown in (a). 
(c) Log-log plot of the surface width vs time for systems of different sizes
evolving at the same diffusion constant $\nu=0.1768$. (d) Time evolution of the effective 
exponent (\ref{eqBeta}) for the data shown in (c). The data in this and the following figures 
have been obtained for $D=1$.}
\label{fig1}
\end{figure}

The stochastic EW equation (\ref{eqEW}),
can be solved exactly to give us for a fixed value of $\nu$ the width\ (squared)
\begin{equation}
W^2\left( t\right) =\frac D{2\nu L}\sum_n\frac{1-e^{-2\nu tk_n^2}}{k_n^2}
\label{eqWsq}
\end{equation}
where $k_n=2\pi n/L$ and the sum is over $\left[ -L/2,L/2\right] $ but {\em %
excluding} the zero mode:{\em \ }$n=0$. 
In Figure \ref{fig1}a,c we show the time dependence of the surface width for,
respectively, the case with fixed $L=1000$ at different $\nu$'s and the case
with a fixed $\nu=0.1768$ and various $L$'s. As for the RDSR process one
distinguishes three regimes separated by two crossover points: a random
deposition (RD) regime, followed by a EW regime, with a final crossover to
the saturation regime. In contrast to Family's original model, the initial RD 
process is not confined to very early times
$t\leq 1$ but might extend to larger times. In fact, the crossover time 
$t_1$ between the RD and the EW regimes is shifted to higher values for
decreasing diffusion constants and diverges in the limit of vanishing $\nu$.
As the crossover is smeared out, we identify the
crossover point with the intersection point of the straight lines fitted to
the two linear regimes in the log-log plots.
We have the identity 
$W^2=t$ in the RD regime, yielding the width $W_1=\sqrt{t_1}$ at the
crossover point. In the EW regime the relation between width and deposition
time changes to $W\propto t^{1/4}$. This regime extends up to a second
crossover point $(t_2,W_2)$, whose precise location depends on the
values of $\nu$ and $D$ and beyond which the final saturation regime prevails. 
The crossover
between the different regimes is further illustrated in Fig. \ref{fig1}b,d
where we show the time evolution of the effective exponent 
\begin{equation}
\beta _{eff}=\frac{d\log W}{d\log t} =\frac{\nu t\sum\limits_n e^{-2\nu tk_n^2}}{\sum\limits_n \left[ 1-e^{-2\nu
tk_n^2}\right] /k_n^2}~.  \label{eqBeta} 
\end{equation}
for the two cases.

We end this Section with a few remarks:

\begin{itemize}
\item  The (``first'') crossover from the RD to the EW regime, denoted by $%
\left( t_1,W_1\right) $, depends only on $\nu$. For the range of $L$'s we
explored, we find $W_1^2=t_1\cong \tau /\nu$, with 
a constant $\tau \cong 0.148$. In the $L\rightarrow
\infty $ limit, $\tau \rightarrow 1/2\pi $ (as shown in the Appendix, see also \cite{Ama92}).

\item  The (``second'') crossover from the EW to the saturation regimes,
denoted by $\left( t_2,W_2\right) $, depends on both $\nu$ and $L$. $W_2$ may
be identified with the saturation width, i.e., $\left( DL/8\pi ^2\nu \right)
\sum_nn^{-2}$ \cite{Ama92}. 
Combining this result with the
line drawn through the EW regime, we arrive at $t_2=\left( L/24\tau \right)
^2t_1$ (see Appendix).
\end{itemize}

%
\section{Sudden change of growth conditions}

With a clear picture of the properties of a surface growing under a {\em %
constant} diffusion constant, we proceed to study the time evolution of
the roughness when $\nu$ is suddenly changed. 
To investigate the response of
the growth process to this change, we use the following protocol: We start
at $t=0$ with a flat surface and let the surface grow at $\nu_i$ until time $%
t=s$, at which point we change the diffusion constant to the final value $\nu_f$. The
change of roughness is then monitored through the time evolution of $W^2$.

As our system displays three different roughness regimes (RD, EW, and
saturation), there can be in principle nine scenarios for the change to be
arranged. They can be distinguished conveniently by (1) $\nu_i$, the diffusion constant
of the initial growth, (2) $\nu_f$, the final value of the diffusion constant, and (3)
$s$, the time at which $\nu$ is suddenly changed.
By choosing these three controls judiciously,
we can access all the scenarios. However, covering all cases in detail is
not the aim of this paper. Instead, we are interested in the new phenomena
associated with the $t$-dependence of the width, $W^2(t,s)$, after
(up- or down-) quenches. Obviously, for $t\gg s$, the surface roughness will
settle into the value in an ``unperturbed'' system, grown at $\nu_f$
from the start. Denoted by $W_u^2(t)$, it is just expression (\ref{eqEW})
with $\nu = \nu _f$. We also refer to this as the
``reference system.'' To highlight the changes, we also study the difference 
(with $t > s$)
\begin{equation}
\Delta W^2(t,s)=|W^2(t,s)-W_u^2(t)|  \label{eq_dW2}
\end{equation}
between quenched system and the reference. Clearly, this quantity reveals
how the roughness of the growing surface adapts itself to the new
``experimental'' condition, and behaves very differently for the various
cases. Our goal is to map out the regions in the $\nu_i$-$\nu_f$-$s$ space
corresponding to the novel behavior following a quench.

To be precise, we will evolve the height $h(x,t)$ starting from $h(x,0)=0$
with $\nu _i$ to time $s$ and then continue
with $\nu _f$ until time $t$. At that point, we compute the width squared
and denote it by $W^2(t,\nu _f;s,\nu _i)$.

Our starting point is the exact solution of (\ref{eqEW})
\begin{eqnarray}
\tilde{h}(k_n,t)& =& e^{-\nu_fk_n^2(t-s)}\, \int\limits_0^s\,dt^{\prime }\,e^{-\nu _ik_n^2(s-t^{\prime
})}\tilde{\eta}(k_n,t^{\prime })\nonumber \\
&& +\int\limits_s^tdt^{\prime \prime }\,e^{-\nu
_fk_n^2(t-t^{\prime \prime })}\tilde{\eta}(k_n,t^{\prime \prime })
\end{eqnarray}
written here in terms of the Fourier amplitudes for $h(x,t)$ and $\eta (x,t)$%
: $\tilde{h}(k_n,t)=\int \, dx\, e^{ik_nx}\, h(x,t)$, etc. Since the noise is
delta-correlated, $\langle | \tilde{h}|^2\rangle $
simplifies so that the width square is
\begin{eqnarray}
W^2(t,\nu _f;s,\nu _i) & = & \frac DL\sum\limits_n\left[ e^{-2\nu
_fk_n^2(t-s)}\int\limits_0^s\,dt^{\prime }\,e^{-2\nu _ik_n^2(s-t^{\prime
})} \right. \nonumber \\
&& \left. +\int\limits_s^tdt^{\prime }\,e^{-2\nu _fk_n^2(t-t^{\prime })}\right] ~.
\label{eqWsq3}
\end{eqnarray}
Since the width square $W_u^2(t,\nu _f)$ of the unperturbed system is given
by
\begin{equation}
W^2(t,\nu _f;s,\nu _i)=\frac DL\sum\limits_n\int\limits_0^tdt^{\prime
}\,e^{-2\nu _fk_n^2(t-t^{\prime })} ~,  \label{eqWsq2}
\end{equation}
we arrive at the exact result for the difference $\Delta W^2= \left| W^2(t,\nu
_f;s,\nu _i)-W_u^2(t,\nu _f) \right| :$
\begin{eqnarray}
\Delta W^2 &=&\left| \frac DL\sum\limits_ne^{-2\nu
_fk_n^2(t-s)}\right. \nonumber \\
&& ~~~~ \left. \int\limits_0^s\,dt^{\prime }\, \left[ e^{-2\nu
_ik_n^2(s-t^{\prime })}-e^{-2\nu _fk_n^2(s-t^{\prime })}\right]\right|   \nonumber
\\
&=&\left| \frac DL\sum\limits_ne^{-2\nu _fk_n^2(t-s)}\right. \nonumber \\
&& ~~~~ \left. \left[ \frac{1-e^{-2\nu
_ik_n^2s}}{2\nu _ik_n^2}-\frac{1-e^{-2\nu _fk_n^2s}}{2\nu _fk_n^2}\right] \right| ~.
\label{eqWsq4}
\end{eqnarray}
For later convenience, we define $\Omega \equiv \nu _f\Delta
W^2(t,\nu _f;s,\nu _i)$ and note that it depends only on {\em three} scaling
variables:
\[
\mu \equiv \nu _i/\nu _f,\quad \sigma \equiv \nu _fs,\quad \rho \equiv
t/s\,\,.
\]
Explicitly, we have
\begin{eqnarray}
\Omega (\mu ,\sigma ,\rho ) &\equiv &\nu _f\Delta W^2 \\
&=&\left| \frac DL\sum\limits_ne^{-2k_n^2\sigma \left( \rho -1\right) }\right. \nonumber \\
&& ~~~~ \left. \left[ \frac{%
1-e^{-2k_n^2\mu \sigma }}{2k_n^2\mu }-\frac{1-e^{-2k_n^2\sigma }}{2k_n^2}%
\right]\right| \,\,.  \label{Omega}
\end{eqnarray}

\begin{figure}[h]
\centerline{\epsfxsize=3.40in\  \epsfbox{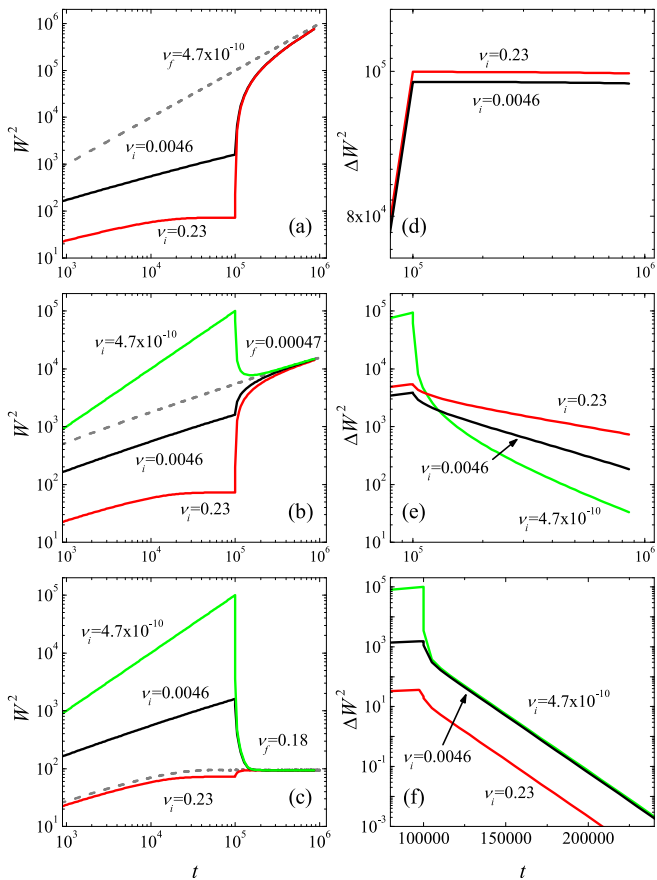}}
\caption{(Color online) (a-c) Time evolution of the width square in case the diffusion constant is
changed after $10^5$ time steps. 
The dashed lines show $W_u^2(t)$ for an
unperturbed surface growing at constant $\nu_f$. In (a) the
quench is to the RD regime, whereas in (b) and (c) the quenches are to the
EW and saturation regimes, respectively. (d-f) The same cases as shown in
(a-c), but now the difference $\Delta W^2$, see Eq. (\ref{eq_dW2}), is
plotted. Qualitative different behavior is observed, depending on the regime
that the unperturbed system has at the quench time. }
\label{fig2}
\end{figure}

To set the stage for discussions, we begin with the data for some typical
cases, all with $s=10^5$, shown in Fig. \ref{fig2}. In order to be able to
discuss the different cases for a fixed $s$, we must work with a relatively
small system: $L=400$. 
The dashed lines in (a-c) represent $W_u^2(t)$ in
an unperturbed system. With $\nu_f=4.7 \cdot 10^{-10}$, $0.00047$, and $0.18$, the surface is, at the
time of the quench, in the (a) RD, (b) EW, and (c) saturation regimes,
respectively. The corresponding differences, $\Delta W^2(t,s)$, are shown in
Fig. \ref{fig2}d-f. 

The effects of two up-quenches into the RD regime, from the EW ($\nu_i=0.0046$)
and the saturation ($\nu_i=0.23$) regimes, are displayed in Fig. \ref{fig2}%
a. As Fig. \ref{fig2}d shows, for quenches into RD, the width $W^2(t,s)$
cannot reach that of the reference system, $W_u^2(t)$. The ``best'' $\Delta
W^2(t,s)$ can achieve is a constant. The physical origin of this behavior
lies in the linear growth of $W^2$ in the RD regime. Thus, for two
unperturbed systems started at different times (say, $t=t_0$ and $t_1$), the
difference $W_0^2-W_1^2$ is just a constant: $t_1-t_0$. In our case, the
correlated growth up to time $s$ endowed our surface with a smaller $%
W^2\left( s\right) $ than the reference $W_u^2(s)$. Immediately after the
quench, correlated growth is simply replaced by independent growth of
different columns and the width $W^2\left( s\right) $ is ``frozen'' in as a
kind of ``initial condition'' (at $t=s$). As a result, the difference $%
\Delta W^2=W_u^2(t)-W^2(t,s)$ remains at the value $W_u^2(s)-W^2\left(
s\right) $. Of course, if we follow these two systems further in time, $%
\Delta W^2$ will eventually vanish. 

Turning next to quenches to the EW regime, we found the most interesting
behavior (Fig. \ref{fig2}b,e). The difference $\Delta W^2$ initially
decreases rapidly, before crossing over to a slower, power-law decay at
larger times ($t\gg s$): 
\begin{equation}
\Delta W^2\sim t^{-\gamma }
\end{equation}
For example, for the cases shown in Fig. \ref{fig2}e, we measure 
close to the end of the time interval the exponents
$\gamma = 0.72$ for $\nu_i = 0.23$, $\gamma = 1.24$ for $\nu_i = 0.0046$,
and $\gamma = 1.65$ for $\nu_i = 4.7 \cdot 10^{-10}$.
As we argue below, these are effective values, the asymptotic values of
$\gamma $ being $1/2$ or $3/2$.
Below we will also discuss in more detail the conditions under which these values
can be expected. Finally, for a quench to the saturation regime, $\Delta W^2$
decays exponentially: 
\begin{equation}
\Delta W^2\sim \exp (-\kappa t)
\end{equation}
with a decay constant $\kappa $ that depends both on the value of the final
diffusion constante $\nu_f$ and on the system size $L$.      

Up to now, we have shown only the simplest situation where the system is
well within a given initial regime at the moment of the quench and, in
addition, that it has time to relax into a well defined regime of the
reference system. Clearly, as we let the final system evolve further, it may
crossover to a different regime (e.g., in case of Fig. \ref{fig2}a,d a crossing
over to the EW and saturation regimes will take place for larger $t$). Therefore, we should expect the
general relaxation process to be quite complex. 

The closed forms (\ref{eqWsq4},\ref{Omega}) are not particularly transparent,
as they involve all possible crossover behaviors. To shed some light on the
various scenarios, we consider some limiting cases where simple properties
(exponentials and powers) can be extracted. A straightforward case is $\nu
_f(t-s)\gg L^2$, so that $\nu _fk_n^2(t-s)$ is always large. Then, the
leading decay is exponential, namely $e^{-8\pi ^2\nu _f(t-s)/L^2}$, since
the other terms in the sum will be much smaller:
\begin{eqnarray}
&& \left( e^{-8\pi ^2\nu _f(t-s)/L^2}\right) ^4,~\left( e^{-8\pi ^2\nu
_f(t-s)/L^2}\right) ^9,\cdots \nonumber \\
&& ~~~~\ll e^{-8\pi ^2\nu _f(t-s)/L^2}~.
\end{eqnarray}
In the opposite limit, where $\nu _fs,\nu _is\ll 1$, we have
\begin{equation}
\left[ \frac{1-e^{-2\nu _ik_n^2s}}{2\nu _ik_n^2}-\frac{1-e^{-2\nu _fk_n^2s}}{%
2\nu _fk_n^2}\right] \cong (\nu _f-\nu _i)s^2k_n^2
\end{equation}
to leading order. The summation over $n$ then yields, for $t-s\gg 1$, a
power-law decay: $t^{-3/2}$, i.e., $\gamma =3/2$.

In order to explore the three-dimensional parameter space $\left( \mu
,\sigma ,\rho \right) $ in a comprehensive way, we evaluate numerically the
closed form (\ref{Omega}). The exponent $\gamma $ can be defined effectively
as $- d\log (\Omega )/d\log (t)$. In Fig. \ref{fig5} we show the contour plot
of $\gamma $ as a function of $\mu $ and $\sigma $ for $\rho =64$. This plot
reveals four different regimes: the regime where $\Omega $ or $\Delta W^2$
is constant (labeled by $\gamma =0$), two power-law regimes with values $%
\gamma =1/2$ and $\gamma =3/2$, and finally a regime of exponential decay
for large $\sigma $. The different regimes are separated by crossover
regions where the effective exponent does not lock-in into one of the values
0, 1/2, or 3/2.

\begin{figure}[t]
\centerline{\epsfxsize=3.40in\  \epsfbox{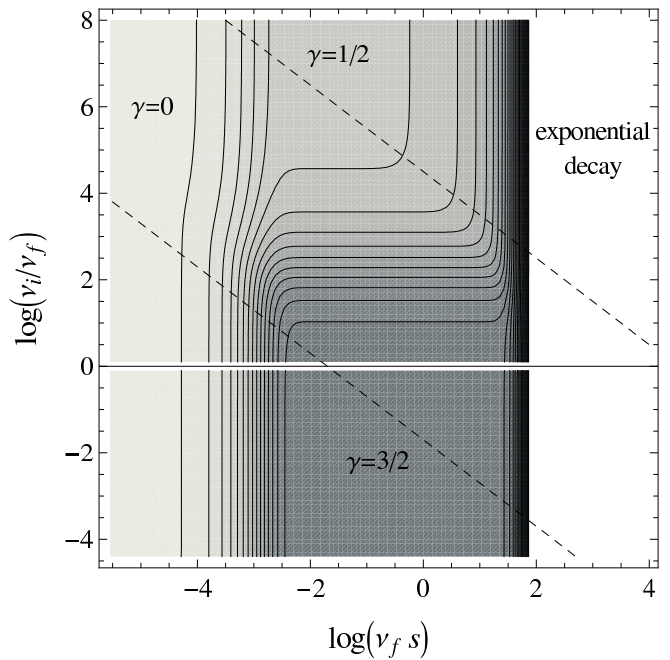}}
\caption{Contour plot of $\gamma $ as a function of $\nu_f s$ and $\nu_i /\nu_f $
for $t/s = \rho =64$. Four different regimes, separated by crossover regions,
are identified. The two dashed lines separate the three qualitatively different
types of behavior encountered when plotting the effective exponent as a
function of $t$, see Fig. \ref{fig5}. }
\label{fig3}
\end{figure}

\begin{figure}[h]
\centerline{\epsfxsize=3.40in\  \epsfbox{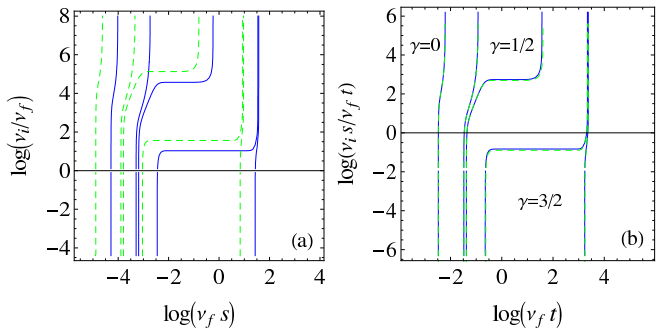}}
\caption{(Color online) (a) Contour plots of $\gamma $ as a function of $\nu_f s=\sigma $ and 
$\nu_i /\nu_f $ for $t/s=\rho =64$ (full lines) and $256$ (dashed 
lines). Only contours bounding the $\gamma =0$, $1/2$ and $3/2$ regimes are
shown. (b) The same contour plots as shown in (a) but as a function of $\nu_f t
$ and ${\nu_i s}/{\nu_f t}$. An approximate collapse of the contours is
observed. }
\label{fig4}
\end{figure}

\begin{figure}[h]
\centerline{\epsfxsize=3.40in\  \epsfbox{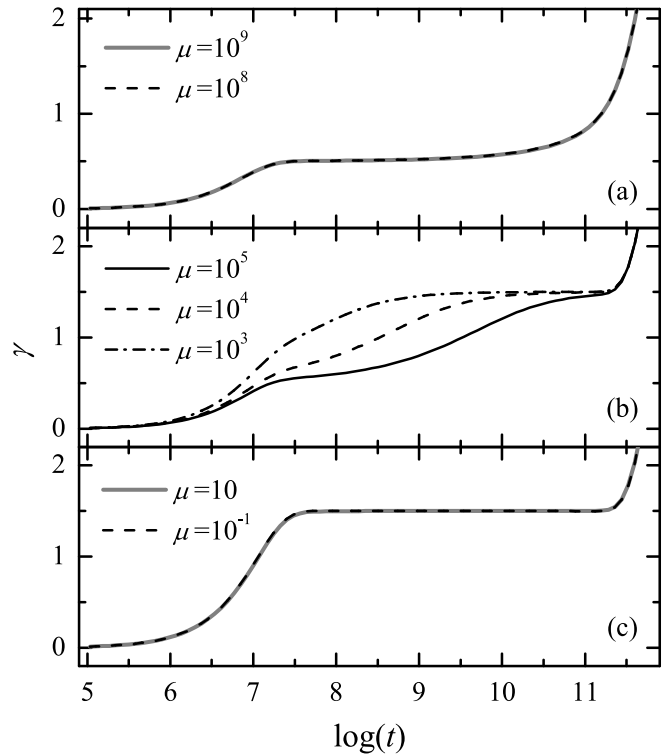}}
\caption{The three different quench types illustrated by the time dependence
of the effective exponent $\gamma $. In all cases the quench takes place at 
$s=10^5$. Type (a) includes quenches with initial conditions $\nu _fs$ and 
$\nu _i/\nu _f$ located above the upper dashed line in Fig \ref{fig4}. For
type (b) the starting point is located between the two dashed lines in Fig 
\ref{fig4}, wheres for type (c) the initial conditions put the starting
point below the lower dashed line. }
\label{fig5}
\end{figure}

Fig. \ref{fig4}a shows how the extensions of the four regimes depend on the
value of $\rho $ in cases where $\rho \gg 1$. Interestingly, an increase of
the value of $\rho $ mainly shifts the contours in the $\log (\sigma )$ {\em %
vs. }$\log (\mu )$ plot along the $(-1,1)$ direction. This is shown in Fig. 
\ref{fig4}b where we plot $\log (\nu _ft)=\log (\sigma \rho )$ {\em vs.} $%
\log (\nu _is/\nu _ft)=\log (\mu /\rho )$. This way of plotting indeed leads
to an approximate data collapse, which gets better for larger values of $%
\rho $. That this data collapse is only approximate also follows from
inspection of the exact solution (\ref{Omega}). Still, Fig. \ref{fig4}b
nicely allows us to visualize the extent of the different dynamic regimes
for large ratios $\rho $.

Finally, in Fig. \ref{fig5}, we discuss the change of the effective exponent 
$\gamma $ as a function of $t$ for various values of $\mu $ (and $s=10^5$).
Note that an increasing time $t$ corresponds in Fig. \ref{fig3}
approximately to a cut along the $(-1,1)$ direction, so that we can
distinguish three typical scenarios, separated by the dashed lines there.
Along the upper dashed line, Fig. \ref{fig5}a shows the effective exponent
rising to the $\gamma =1/2$ plateau where it remains for a long time before
crossing over to the regime where the difference $\Delta W^2$ vanishes
exponentially fast (``$\gamma =\infty $''). For the region above this line
in Fig. \ref{fig3}, we can expect similar results. Along the lower dashed
line, the same behavior is seen, except that the plateau value is now $%
\gamma =3/2$ (Fig. \ref{fig5}c). This can also be expected for the region
below this lower dashed line in Fig. \ref{fig3}. Between these two protocols,
a more complex behavior is encountered, as the effective exponent shows
some tendency to lock-in at both values 1/2 and 3/2 (see Fig. \ref{fig5}b).
We should remind the reader that the $\gamma \left( t\right) $ curves shown
here are applicable for all varieties of quenches (different quench times,
as well as different initial and final values of $\nu$) as long as the rescaled time $\nu
_fs$ is fixed.

It is interesting to note that a change of the noise during the growth process also yields
a power-law relaxation, as discussed in \cite{Maj96}. Thus for the one-dimensional EW equation it was
observed that $\Delta W^2 \sim t^{-1/2}$ if the system is in the EW regime before and after
changing the noise of the system. The situation studied in \cite{Maj96} is therefore comparable 
to our case, even though we do observe a richer behavior when changing the diffusion constant.

\section{Discussion and outlook}

The morphology and roughness of growing surfaces depend in a crucial way on
experimental conditions. Motivated by various observations of a change of
the roughness universality class when changing experimental conditions, we
have studied in this work how a growth process reacts to a sudden change of
the diffusion constant. 
Exploiting the fact that the stochastic EW
equation can be solved exactly, we carried out a comprehensive study of the
response of the growing interface. 
Four main relaxation regimes, separated by
crossovers, were identified. For extremely long times any finite system will
eventually relax in an exponential way, due to the presence of the
saturation regime, but this only takes place after an earlier power-law
relaxation. For finite times, a growing surface rapidly reacts to a change of 
$\nu$ such that its morphology (roughness) approaches that of a reference
system that was allowed to grow under stable external conditions. If the
change of growth conditions takes place at a time where both the quenched and the
reference system are in a correlated growth regime, than the relaxation
process is governed by power-laws over many time decades. This is our main
result (see also \cite{Maj96}), and we expect this to hold true for other systems.

Of course, we do not expect that our results for a one-dimensional system
can be applied quantitatively to any of the experimental systems in which a
temperature dependence of the roughness exponent has been observed.
Nevertheless, we expect that these results should be generic
for growth processes with a sudden change of external conditions, and that
the intriguing signatures revealed in our study can be generalized to more
realistic models, so that they can be observed in experiments on physical
systems.

{}From a more theoretical point of view, our study reveals that 
growth processes provide one of the rare cases where a power-law relaxation
is generically observed when quenching from one correlated state to another. More specifically,
the exactly solvable Edwards-Wilkinson equation allows us to derive analytical expressions,
thus permitting a complete investigation of the various possible scenarios. 
It is to be expected that the power-law relaxation encountered at a quench also entails interesting 
aging processes not studied in the past. Indeed, in the few published studies of aging 
in growth processes \cite{Rot06,Bus07,Bus07a} constant model parameters
were always assumed. 

Our study can be extended in various directions. On the one hand, we can
study quenches in systems where the interface is stabilized not by surface
tension, but by a curvature Hamiltonian. The simplest case is given by the
noisy Mullins-Herring equation \cite{Mul63,Wol90}. As this is again a linear
stochastic differential equation, we can follow the same strategy as in the
present work and investigate the response to a temperature quench by
analyzing exact expressions. On the other hand, we can also extend our study
to systems that are of direct relevance for thin film growth: the
Kardar-Parisi-Zhang (KPZ) \cite{Kar86} and the conserved KPZ universality
classes \cite{Wol90,Sar91}. As exact expressions for these non-linear
systems are not available, we plan to integrate these equations numerically
and to simulate quenches in microscopic models belonging to the same
universality classes.

\begin{acknowledgments}
This work was supported in part by the US National
Science Foundation through DMR- 0705152 (R.K.P. Zia) and
DMR-0904999 (M. Pleimling). 
\end{acknowledgments}

\section*{Appendix}
Here, we show how to compute the first crossover time $t_1$ in the limit of
infinite $L$ (see also \cite{Ama92}). Since this point is defined as the intersection of the RD
regime ($W^2=Dt$) and the EW regime ($W^2\cong At^{1/2}$), we have 
\begin{equation}
t_1=(A/D)^2
\end{equation}
so that the problem reduces to finding the amplitude $A$ associated with EW
growth. For $L\rightarrow \infty $ , the sum in expression (\ref{eqWsq}) can
be replaced by an integral, which can be computed to extract $A$. Further
simplification occurs if we consider $\partial _tW^2=D\int_{-\pi }^\pi
e^{-2\nu t\theta ^2}d\theta /2\pi $ instead. Imposing the ansatz $W^2\cong
At^{1/2}$, we arrive at 
\begin{equation}
At^{-1/2}\cong \frac D\pi \int_{-\pi }^\pi e^{-2\nu t\theta ^2}d\theta .
\label{A1}
\end{equation}
Transforming to $\xi \equiv \sqrt{2\nu t}\theta $, we have 
\begin{equation}
A\cong \frac D{\sqrt{2\nu }\pi }\int_{-\pi \sqrt{2\nu t}}^{\pi \sqrt{2\nu t}%
}e^{-\xi ^2}d\xi  .
\end{equation}
So, as $t\rightarrow \infty $ (or, to be precise, $L\gg t\gg \nu ^{-1}$), we
arrive at $W^2\left( t\right) =At^{1/2}$ as well as 
\begin{equation}
t_1=\left(\frac AD\right)^2=\frac 1{2\pi \nu }= \frac{\tau}{\nu}  \label{exact t1}
\end{equation}
with $\tau = 1/2 \pi$.

Of course, this approach can also be used to extract $t_2$. Equating this $%
W^2$ (i.e., $At^{1/2}$) to the saturation $W^2$ (i.e., $\left( DL/8\pi ^2\nu
\right) \sum_nn^{-2}=DL/24\nu $), we arrive at 
\begin{displaymath}
t_2=\left( \frac {DL}{24\nu A}\right) ^2=\frac \pi {288\nu }L^2\,\,. 
\end{displaymath}
This provides 
\begin{displaymath}
\frac{t_2}{t_1}=\left( \frac{\pi L}{12}\right) ^2  = \left( \frac{L}{24 \, \tau} \right)^2
\end{displaymath}
and 
\begin{displaymath}
\frac 1\lambda =\,\,\log \sqrt{\pi L/12}. 
\end{displaymath}
We should emphasize again that these results are exact in the $L\rightarrow
\infty $ limit. 

\end{document}